\documentclass[english]{article}
\usepackage{fontenc}
\usepackage[latin1]{inputenc}
\usepackage{geometry}
\usepackage{amsmath}
\usepackage{fontenc}
\usepackage{fontenc}
\usepackage{graphicx}
\usepackage{babel}

\input{tcilatex}

\begin{document}

\title{\textbf{Comments on spin operators and spin-polarization states of }$%
\mathbf{2+1}$\textbf{\ fermions}}
\author{S.P. Gavrilov\thanks{%
Dept. Física e Química, UNESP, Campus de Guaratinguetá, Brazil; Permanent
address: Dept. of General and Experimental Physics, Herzen State Pedagogical
University, St.-Petersburg, Russia; email: gavrilovsergeyp@yahoo.com}, D.M.
Gitman\thanks{%
E-mail: gitman@dfn.if.usp.br}, and J.L. Tomazelli\thanks{%
Dept. Física e Química, UNESP, Campus de Guaratinguetá, Brazil} \\
\\
Instituto de Física, Universidade de São Paulo,\\
Caixa Postal 66318, CEP 05315-970 São Paulo, SP, Brazil}
\maketitle

\begin{abstract}
In this brief article we discuss spin polarization operators and spin
polarization states of $2+1$ massive Dirac fermions and find a convenient
representation by the help of $4$-spinors for their description. We stress
that in particular the use of such a representation allows us to introduce
the conserved covariant spin operator in the $2+1$ field theory. Another
advantage of this representation is related to the pseudoclassical limit of
the theory. Indeed, quantization of the pseudoclassical model of a spinning
particle in $2+1$ dimensions leads to the $4$-spinor representation as the
adequate realization of the operator algebra, where the corresponding
operator of a first-class constraint, which cannot be gauged out by imposing
the gauge condition, is just the covariant operator previously introduced in
the quantum theory.
\end{abstract}

I. The $2+1$ spinor field theory \cite{JTD} has attracted in recent years
great attention due to various reasons, e.g., because of nontrivial
topological properties, and due to a possibility of the existence of
particles with fractional spins and exotic statistics (anyons), having
probably applications to fractional Hall effect, high-$T_{c}$
superconductivity and so on \cite{Wil}. In many practical situations the
quantum behavior of spin $1/2$ fermions (from now on simply called fermions)
in $2+1$ dimensions can be described by the corresponding Dirac equation
with an external electromagnetic field. The main difference between the
relativistic quantum mechanics of fermions in $3+1$ and in $2+1$ dimensions
is related to the different description of spin polarization states. It is
well know that in $3+1$ dim. there exist two massive spin $1/2$ fermions,
the electron and its corresponding antiparticle, i.e., the positron. Both
the electron and the positron have two spin polarization states. In $2+1$
dim. there exist four massive fermions: two different types of electrons and
two corresponding positrons. In contrast to the situation in $3+1$ dim.,
each particle in $2+1$ dim. has only one polarization state. We recall that
constructing the covariant and conserved spin operators for the $3+1$ Dirac
equation in an external field is an important problem as regards finding
exact solutions of this equation and specificating the spin polarization
states \cite{BagGi90}. Here, there do not exist universal covariant
conserved spin operators which relate to any external field, for each
specific configuration of the external field one has to determine such
operators \cite{ST68}. At first glance, the problem does not exist in $2+1$
dim., since each fermion has only one spin polarization state. Nevertheless,
the spin (or spin magnetic momentum) as a physical quantity in $2+1$ dim.
does exist, and therefore, the corresponding operators do exist. One can
see, by solving the Dirac equation in $2+1$ dim., that knowledge of such
spin operators is very useful for finding physically meaningful solutions.
Moreover, it turns out that in $2+1$ dim. the appropriate spin operator
serves at the same time as a particle species operator whilst its explicit
expression is useful for interpretation of the theoretical constructions. In
this brief article, we discuss spin-polarization operators and
spin-polarization states of $2+1$ massive Dirac fermions and some convenient
representations for their description.

II. It is well known that in $2+1$ dim. (as well as in any odd dimensions)
there exist two inequivalent sets (representations) of gamma-matrices. In
fact, the proper orthocronous Lorentz group $L_{+}^{\uparrow }$, in a
pseudo-euclidean space $\mathcal{M}$, can be identified with the $SL(2,R)$
group of real unimodular $2\times 2$ matrices, associated to linear
transformations of unity determinant in a real two-dimensional vector space %
\cite{Bin}. Considering the space $\mathcal{M}$ of real hermitian matrices
spanned by the vector basis $\{\tau _{\alpha }\}$, 
\begin{equation*}
\tau _{0}=\mathbf{1}\,\,,\,\,\tau _{1}=\sigma _{3}\,\,,\,\,\tau _{2}=\sigma
_{1}\,\,,
\end{equation*}%
where $\sigma _{1,3}$ belong to the set of Pauli matrices $\sigma
_{i}\,,\,i=1,2,3$, one can associate to each vector in this space a matrix
in $\mathcal{M}$ via the isomorphism 
\begin{equation*}
L_{+}^{\uparrow }\approx SL(2,R)/Z,
\end{equation*}%
where Z denotes the center $\{I,-I\}$ of the $SL(2,R)$ group, which
constitutes the (real) spinor representation of the Lorentz group.

An operator in $SL(2,R)$ can be represented by a matrix in the hermitian
basis $\{\tau _{\alpha }\}$ or simply by a matrix formed by the product of
any two elements of such a basis as, for example, the anti-hermitian matrix $%
\tau _{3}=\tau _{1}\tau _{2}$ in the $SL(2,R)$ associated algebra. The new
(non-hermitian) basis $\{\tau _{1},\tau _{2},\tau _{3}\}$ of $SL(2,R)$ is
also the set of generators of the real Clifford algebra \cite{Coq} 
\begin{equation*}
\lbrack \tau _{i},\tau _{j}]_{+}=2\overline{g}_{i\,j}\,\,;\,\,i,j=1,2,3\,\,,
\end{equation*}%
where $\overline{g}$ is the metric tensor of a pseudo-euclidean space of
signature $(+,+,-)$.

The complexifications 
\begin{equation}
\Gamma _{s}^{0}=\tau _{1}=\sigma _{3}\,\,,\,\,\Gamma _{s}^{1}=\tau
_{3}=i\sigma _{2}\,\,,\,\,\Gamma _{s}^{2}=-si\tau _{2}=-si\sigma
_{1},\;s=\pm 1\,\,,  \label{1}
\end{equation}%
of the above Clifford algebra give rise to the algebra of different
representations for Dirac gamma-matrices, labeled by the subscript $s=\pm 1$.

As a consequence, there exist, respectively, two different Dirac equations
and two different Lagrangians for the corresponding spinor fields. If an
external electromagnetic field is present, then the particle ($\zeta =1$)
and antiparticle ($\zeta =-1$) with the charges $\zeta e,\;e>0$ respectively
obey the Dirac equations in which the operator $i\partial _{\mu }$ has to be
replaced by $P_{\mu }=i\partial _{\mu }-\zeta eA_{\mu }\left( x\right) ,$
where $A_{\mu }\left( x\right) $ are electromagnetic potentials. Thus, in
fact, in $2+1$ dim. we have four massive fermions (let us call further the
two different types of fermions up and down particles) and respectively four
types of solutions of the $2+1$ Dirac equation ($2$-spinors $\Psi ^{\left(
\zeta ,s\right) }\left( x\right) $): 
\begin{align}
& \left( \Gamma _{s}^{\mu }P_{\mu }-m\right) \Psi ^{\left( \zeta ,s\right)
}\left( x\right) =0\,,\;x=\left( x^{\mu }\right) \,,\;\mu =0,1,2\,,  \notag
\\
& P_{\mu }=i\partial _{\mu }-\zeta eA_{\mu }\left( x\right) \,,\;s,\zeta
=\pm 1\,.  \label{2}
\end{align}%
In such a picture (and in stationary external fields that do not violate the
vacuum stability), the only physical states are those from the upper energy
branch, and only such states can be used in second quantization \cite%
{FreGaG04}.

III. In order to define a spin magnetic momentum of the $2+1$ massive
fermions let us set the external field to be a uniform constant magnetic
field. In $2+1$ dim., the magnetic field has only one component $%
F_{21}=-F_{12}=B=\mathrm{const}$. The sign of $B$ defines the ''direction''
of the field, the positive $B$ corresponds to the ''up'' direction whereas
the negative $B$ corresponds to the ''down'' direction. In such a
background, the equation (\ref{2}) can be reduced to the stationary form 
\begin{align}
& H^{\left( \zeta,s\right) }\Psi_{n}^{\left( \zeta,s\right) }\left( \mathbf{x%
}\right) =\varepsilon_{n}^{\left( \zeta,s\right) }\Psi _{n}^{\left(
\zeta,s\right) }\left( \mathbf{x}\right) \,,\;\;H^{\left( \zeta,s\right)
}=-\Gamma_{s}^{0}\Gamma_{s}^{k}P_{k}+\Gamma_{s}^{0}m\,,  \notag \\
& \Psi^{\left( \zeta,s\right) }\left( x\right) =\exp\left(
-i\varepsilon^{\left( \zeta,s\right) }x^{0}\right) \Psi^{\left(
\zeta,s\right) }\left( \mathbf{x}\right) \,,\;\varepsilon_{n}^{\left(
\zeta,s\right) }>0\,,\;\mathbf{x}=\left( x^{1},x^{2}\right) \,.  \label{3}
\end{align}
As usual, we pass to the squared equation through the ansatz 
\begin{equation}
\Psi^{\left( \zeta,s\right) }\left( \mathbf{x}\right) =\left[ \Gamma
_{s}^{0}\varepsilon+\Gamma_{s}^{k}P_{k}+m\right] \Phi^{\left( \zeta
,s\right) }\left( \mathbf{x}\right)  \label{5}
\end{equation}
to obtain the following equation 
\begin{align}
& \left[ \varepsilon_{n}^{2}-D^{\left( \zeta,s\right) }\right] \Phi
_{n}^{\left( \zeta,s\right) }\left( \mathbf{x}\right) =0\,,  \notag \\
& D^{\left( \zeta,s\right) }=m^{2}+\mathbf{P}^{2}+\frac{i}{4}\zeta
eF_{\mu\nu}\left[ \Gamma_{s}^{\mu},\Gamma_{s}^{\nu}\right] =m^{2}+\mathbf{P}%
^{2}-s\zeta eB\sigma^{3}\,,\;\mathbf{P=}\left( P^{1},P^{2}\right) \,.
\label{4}
\end{align}
The $2$-component spinors $\Phi_{n}^{(\zeta,s)}\left( \mathbf{x}\right) $
may be chosen in the form $\Phi^{\left( \zeta,s\right) }\left( \mathbf{x}%
\right) =f_{n}^{(\zeta,s)}\left( \mathbf{x}\right) \upsilon\,,$ where $%
f_{n}^{(\zeta,s)}\left( \mathbf{x}\right) $ are some functions and $\upsilon$
some constant $2$-component spinors that classify spin polarization states.
We select $\upsilon$ to obey the equation $\sigma^{3}\upsilon =\upsilon$.
One can see that selecting $\upsilon$ to be the eigenvector of $\sigma^{3}$
with the eigenvalue $-1,$ we do not obtain new linearly independent spinors $%
\Psi_{n}^{\left( \zeta,s\right) }\left( \mathbf{x}\right) $. This is a
reflection of the well known fact (see e.g. \cite{JacN91}) that massive $2+1$
Dirac fermions have only one spin polarization state. This reflects the fact
that, in $2+1$ dim., the mass terms in the corresponding Lagrangians for the
spinor fields $\Psi^{\left( \zeta,s\right) }$ are not invariant under parity
transformation, which consists in the inversion of one of the space
coordinate axis, say, the $x$-axis.

In fact, under the transformation 
\begin{equation*}
\mathcal{P}\,:\,\mathbf{x}\rightarrow \mathbf{x}^{\prime }=(-x,y)\,\,,
\end{equation*}%
the spinor $\Psi ^{\left( \zeta ,+\right) }(x)$, which satisfies the above
Dirac equation for fermions of mass $m$, transforms as 
\begin{equation*}
\Psi ^{\left( \zeta ,+\right) }(x)\rightarrow \Psi ^{\prime }{}^{\left(
\zeta ,-\right) }(x^{\prime })=\mathcal{P}\Psi ^{\left( \zeta ,+\right)
}(x)\,\,,
\end{equation*}%
where the components of the spinor $\Psi ^{\prime }{}^{\left( \zeta
,-\right) }$ obey the equations%
\begin{eqnarray*}
(P_{0}\Psi ^{\prime }{}_{1}^{\left( \zeta ,-\right) }(x^{\prime })+P_{1}\Psi
^{\prime }{}_{2}^{\left( \zeta ,-\right) }(x^{\prime })+iP_{2}\Psi ^{\prime
}{}_{2}^{\left( \zeta ,-\right) }(x^{\prime })) &=&m\Psi ^{\prime
}{}_{1}^{\left( \zeta ,-\right) }(x^{\prime }), \\
(-P_{0}\Psi ^{\prime }{}_{2}^{\left( \zeta ,-\right) }(x^{\prime
})-P_{1}\Psi ^{\prime }{}_{1}^{\left( \zeta ,-\right) }(x^{\prime
})+iP_{2}\Psi ^{\prime }{}_{1}^{\left( \zeta ,+\right) }(x^{\prime }))
&=&m\Psi ^{\prime }{}_{2}^{\left( \zeta ,+\right) }(x^{\prime }),
\end{eqnarray*}
in any Lorentz reference frame. Hence, we can verify that for 
\begin{equation*}
\Psi ^{\prime }{}_{1}^{\left( \zeta ,-\right) }(x^{\prime })=\Psi
_{2}^{\left( \zeta ,+\right) }(x)\,\,\,\,\text{\textrm{and}}\,\,\,\,\Psi
^{\prime }{}_{2}^{\left( \zeta ,-\right) }(x^{\prime })=-\Psi _{1}^{\left(
\zeta ,+\right) }(x)\,\,,
\end{equation*}%
i.e., 
\begin{equation*}
\Psi ^{\prime }{}^{\left( \zeta ,-\right) }(x^{\prime })=\Gamma _{+}^{1}\Psi
^{\left( \zeta ,+\right) }(x)\,\,,
\end{equation*}%
we obtain a Dirac equation for a particle of mass $-m$. Therefore, we
associate to the operator $\mathcal{P}$ which acts on $\Psi ^{\left( \zeta
,+\right) }(x)$ the gamma-matrix $\Gamma _{+}^{1}$. We are thus led to the
conclusion that in order to get new solutions of Dirac equation from the
corresponding $\Psi ^{\left( \zeta ,s\right) }(x)$ ones, besides an internal
transformation such as the above parity transformation, it is necessary to
perform the change $m\leftrightarrow -m$; these solutions must indeed be
related to the solutions $\Psi ^{\left( \zeta ,-s\right) }(x)$, due to the
existence of only two inequivalent representations of Dirac gamma-matrices.

In a weak magnetic field it follows from (\ref{4}): 
\begin{equation}
\varepsilon_{n}^{\left( \zeta,s\right) }=\left. \varepsilon_{n}^{\left(
\zeta,s\right) }\right| _{B=0}-\mu^{\left( \zeta,s\right) }B\,,\;\;\mu
^{\left( \zeta,s\right) }=\frac{s\zeta e}{2\sqrt{m^{2}+\left(
f_{n}^{(\zeta,s)}\right) ^{-1}\mathbf{P}^{2}f_{n}^{(\zeta,s)}}}\,.  \label{7}
\end{equation}
We have to interpret $\mu^{\left( \zeta,s\right) }$ as the spin magnetic
momentum of $2+1$ fermions. Thus in $2+1$ dim., we have 
\begin{equation}
\mathrm{sign\,}\mu^{\left( \zeta,s\right) }=s\zeta\,.  \label{8}
\end{equation}

One ought to remark that this result matches with the conventional
description of spin polarization in $2+1$ dimensions. Considering the total
angular momentum in the rest frame (see, for example, \cite{JacN91,GitSh}),
one can define the operators $S_{0}^{(s)}$ of spin projection on the $x^{0}$%
-axis, 
\begin{equation}
S_{0}^{(s)}=\frac{i}{4}\left[ \Gamma _{s}^{1},\Gamma _{s}^{2}\right] =\frac{s%
}{2}\sigma ^{3}\,.  \label{9}
\end{equation}%
In the non-relativistic limit we obtain from (\ref{5}) and (\ref{7}), 
\begin{equation*}
\mu ^{\left( \zeta ,s\right) }=\frac{s\zeta e}{2m}\,,\;\Psi _{n}^{(\zeta
,s)}\left( \mathbf{x}\right) =2m\Phi _{n}^{(\zeta ,s)}\left( \mathbf{x}%
\right) .
\end{equation*}%
In such a limit the Dirac spinors $\Psi _{n}^{(\zeta ,s)}\left( \mathbf{x}%
\right) $ are eigenfunctions of the operators (\ref{9}), 
\begin{equation*}
S_{0}^{(s)}\Psi _{n}^{(\zeta ,s)}\left( \mathbf{x}\right) =\frac{s}{2}\Psi
_{n}^{(\zeta ,s)}\left( \mathbf{x}\right) \,.
\end{equation*}%
Thus, one can consider 
\begin{equation}
M^{\left( \zeta ,s\right) }=\frac{\zeta e}{m}S_{0}^{(s)}\,  \label{10}
\end{equation}%
as the spin magnetic momentum operator. However, the operators $S_{0}^{(s)}$%
are not covariant and are not conserved in the external field. Below we
represent a conserved and covariant spin operator for $2+1$ massive fermions.

IV. Let us use a $4$-component spinor representation for the wave functions
to describe particles in $2+1$ dimensions. Namely, let us introduce $4$%
-component spinors of the form 
\begin{equation}
\psi ^{(\zeta ,+1)}\left( x\right) =\left( 
\begin{array}{c}
\Psi ^{(\zeta ,+1)}\left( x\right)  \\ 
0%
\end{array}%
\right) ,\;\;\psi ^{(\zeta ,-1)}\left( x\right) =\left( 
\begin{array}{c}
0 \\ 
\sigma ^{1}\Psi ^{(\zeta ,-1)}\left( x\right) 
\end{array}%
\right) .  \label{13}
\end{equation}%
These $4$-component spinors are representatives of $2$-component spinors $%
\Psi ^{(\zeta ,+1)}\left( x\right) $ and $\Psi ^{(\zeta ,-1)}\left( x\right)
.$ At the same time it is convenient to use three $4\times 4$ matrices $%
\gamma ^{0},\gamma ^{1},$ and $\gamma ^{2}$ taken from the following
representation \cite{AMW89} of $3+1$ gamma-matrices 
\begin{align}
\gamma ^{0}& =\left( 
\begin{array}{cc}
\Gamma _{+1}^{0} & 0 \\ 
0 & -\Gamma _{-1}^{0}%
\end{array}%
\right) ,\;\gamma ^{1}=\left( 
\begin{array}{cc}
\Gamma _{+1}^{1} & 0 \\ 
0 & -\Gamma _{-1}^{1}%
\end{array}%
\right) \,,  \notag \\
\gamma ^{2}& =\left( 
\begin{array}{cc}
\Gamma _{+1}^{2} & 0 \\ 
0 & \Gamma _{-1}^{2}%
\end{array}%
\right) ,\;\gamma ^{3}=\left( 
\begin{array}{cc}
0 & I \\ 
-I & 0%
\end{array}%
\right) \,.  \label{14}
\end{align}%
In the new representation, the $4$-component spinors (\ref{13}) obey the
Dirac equation of the following form 
\begin{equation}
\left( \gamma ^{\mu }P_{\mu }-m\right) \psi \left( x\right) =0\,,\;P_{\mu
}=i\partial _{\mu }-\zeta eA_{\mu }\left( x\right) \,,\;x=\left( x^{\mu
}\right) \,,\;\mu =0,1,2\,.  \label{18}
\end{equation}%
In fact, this equation can be considered as a result of a partial
dimensional reduction of the $3+1$ Dirac equation. Stationary solutions of
equation (\ref{18}) can be expressed via solutions $\Phi _{n}^{\left( \zeta
,s\right) }\left( \mathbf{x}\right) $ of equation (\ref{4}) as follows 
\begin{align}
& \psi _{n}^{\left( \zeta ,s\right) }\left( x\right) =\exp \left(
-i\varepsilon _{n}^{\left( \zeta ,s\right) }x^{0}\right) \left[ \gamma
^{0}\varepsilon _{n}^{\left( \zeta ,s\right) }+\gamma ^{k}P_{k}+m\right]
\varphi ^{\left( \zeta ,s\right) }\left( \mathbf{x}\right) \,,\;\mathbf{x}%
=\left( x^{1},x^{2}\right) \,,  \notag \\
& \varphi ^{\left( \zeta ,+1\right) }=\left( 
\begin{array}{c}
\Phi ^{\left( \zeta ,+1\right) } \\ 
0%
\end{array}%
\right) ,\;\varphi ^{\left( \zeta ,-1\right) }=\left( 
\begin{array}{c}
0 \\ 
\sigma ^{1}\Phi ^{\left( \zeta ,-1\right) }%
\end{array}%
\right) ,\;\varepsilon _{n}^{\left( \zeta ,s\right) }>0\,,  \label{16}
\end{align}%
whereas the energy spectrum is the same as for equation (\ref{4}). One can
easily see that the $4$-spinors $\varphi ^{\left( \zeta ,s\right) }$ are
eigenvectors of the operator $\Sigma ^{3}$ with the eigenvalues $s$ being
the particle \thinspace species 
\begin{equation}
\Sigma ^{3}\varphi ^{\left( \zeta ,s\right) }=s\varphi ^{\left( \zeta
,s\right) }\,,\;\Sigma ^{3}=i\gamma ^{1}\gamma ^{2}=\left( 
\begin{array}{cc}
\sigma ^{3} & 0 \\ 
0 & \sigma ^{3}%
\end{array}%
\right) .  \label{17}
\end{equation}%
The operator $\Sigma ^{3}$ commutes with the squared Dirac equation. This
fact allows us to find a spin integral of motion for the Dirac equation (\ref%
{18}). Such an integral of motion reads 
\begin{equation}
\Lambda =\frac{\mathcal{H}\Sigma ^{3}+\Sigma ^{3}\mathcal{H}}{4m}\,,\;%
\mathcal{H}=-\gamma ^{0}\gamma ^{k}P_{k}+\gamma ^{0}m\,.  \label{19}
\end{equation}%
In the case under consideration, we obtain 
\begin{equation}
\Lambda \psi ^{\left( \zeta ,s\right) }=\frac{s}{2}\psi ^{\left( \zeta
,s\right) }\,,\;\;\Lambda =\frac{1}{2}\gamma ^{0}\Sigma ^{3}=\frac{1}{2}%
\left( 
\begin{array}{cc}
I & 0 \\ 
0 & -I%
\end{array}%
\right) .  \label{20}
\end{equation}

V. Now we can consider $2+1$ QFT of the spinor field that obeys equation (%
\ref{18}). Such a QFT can be obtained by a standard quantization of the
corresponding Lagrangian. Here the field operators have the form 
\begin{equation}
\hat{\psi}\left( x\right) =\left( 
\begin{array}{c}
\hat{\Psi}_{+1}\left( x\right) \\ 
\sigma^{1}\hat{\Psi}_{-1}\left( x\right)%
\end{array}
\right) ,  \label{21}
\end{equation}
where the $2$-component operators $\hat{\Psi}_{s}\left( x\right) $ describe
particles of the $s$-\thinspace species. Decomposing the field (\ref{21})
into the solutions (\ref{16}), we obtain four type creation and annihilation
operators: $a_{s,n}\,$ and $a_{s,n}^{+}\,$\ which are operators of particles
($\zeta=+1$) and $b_{s,n}$ and $b_{s,n}^{+}\,$ which are operators of
antiparticles ($\zeta=-1$). Thus, in the QFT under consideration all the
types of $2+1$ fermions appear at the same footing.

In the QFT one can define the second-quantized operator $\hat{\Lambda}$ that
corresponds to the operator $\Lambda$ of the field theory, 
\begin{equation}
\hat{\Lambda}=\frac{e}{m}\int\hat{\psi}^{\dagger}\Lambda\hat{\psi }d\mathbf{%
x\,.}  \label{24}
\end{equation}
It is easily to verify that such an operator is a scalar under $2+1$ Lorentz
transformations and is conserved in any external field. We call the operator 
$\hat{\Lambda}$ the spin magnetic polarization operator. One can easily see
that this operator is expressed via charge operators $\hat{Q}_{s}$ of $2+1$
fermions as follows: 
\begin{equation}
\hat{\Lambda}=\frac{1}{2m}\left( \hat{Q}_{+1}-\hat{Q}_{-1}\right) \mathbf{,}
\label{22}
\end{equation}
where 
\begin{equation}
\hat{Q}_{s}=\frac{e}{2}\int\left[ \hat{\Psi}_{s}^{\dagger},\hat{\Psi}_{s}%
\right] d\mathbf{x}=e\sum_{n}\left(
a_{s,n}^{+}a_{s,n}-b_{s,n}^{+}b_{s,n}\,\right) \,,\;s=\pm1\,.  \label{23}
\end{equation}
Remark that the eigenvalues of the operator $\hat{\Lambda}$ in the
one-particle sector coincide with the spin magnetic momenta $\mu^{\left(
\zeta,s\right) }=s\zeta e/2m$ of the $2+1$ fermions in the rest frame.

We stress that in particular the use of a spinor representation with more
than $2$-components allows us to introduce the conserved covariant spin
operator in the $2+1$ field theory. There is another argument (which is
related to the first quantization procedure) in favour of such
representations, discussed below.

VI. It was demonstrated in \cite{FreGaG04} that relativistic quantum
mechanics of all the massive $2+1$ fermions can be obtained in course of the
first quantization of a corresponding pseudoclassical action where the
particle \thinspace species $s$ is not fixed. General state vectors are $16$%
-component columns. The states with a definite charge sign $\zeta$ can be
described by $8$-component columns $\phi_{\zeta}.$ The operators of space
coordinates $\hat{X}^{k}$ and momenta $\mathcal{\hat{P}}_{k}$ act on these
columns as%
\begin{equation*}
\hat{X}^{k}=x^{k}\mathbf{I}\,,\;\mathcal{\hat{P}}_{k}=\hat{p}_{k}\mathbf{I}%
\,,\;\hat{p}_{k}=-i\partial_{k}\,.
\end{equation*}
Here, $\mathbf{I}$ is the $8\times8$ unit matrix. Besides the spin degrees
of freedom are related to the operators 
\begin{equation*}
\hat{\xi}^{1}=\frac{i}{2}\mathrm{antidiag}\left( \gamma^{1},\gamma
^{1}\right) \,,\;\hat{\xi}^{2}=\frac{i}{2}\mathrm{diag}\left( \gamma
^{2},\gamma^{2}\right) \,.
\end{equation*}
The operator of a conserved first-class (ungauged) constraint has a form 
\begin{equation*}
\,\,\hat{t}=\hat{\theta}-\hat{S}\,,\;\;\hat{\theta}=\mathrm{diag}\left(
\Lambda,\Lambda\right) \,,\;\hat{S}=2i\hat{\xi}^{2}\hat{\xi}^{1}\,.
\end{equation*}
To fix the gauge at the quantum level, one imposes according to Dirac the
condition $\hat{t}\phi_{\zeta}=0\,$on physical state vectors. At the same
time we chose $\phi_{\zeta}$ to be eigenvectors of the matrix $\hat{\theta} $%
, 
\begin{equation*}
\hat{\theta}\phi_{\zeta,s}=\frac{s}{2}\phi_{\zeta,s}\,.
\end{equation*}
We see that in the first quantized theory under consideration the operator $%
\hat{S}$ acts as the operator $\Lambda$ in the quantum mechanics of item IV,%
\begin{equation*}
\hat{S}\phi_{\zeta,s}=\frac{s}{2}\phi_{\zeta,s}\,.
\end{equation*}
Thus, we can interpret the operator $\hat{S}$ as spin operator.

Finally, there exists a relation between the representations of one-particle
quantum states in terms of $\phi_{\zeta,s}$ and $\psi^{(\zeta,s)}.$ Such a
relation reads: 
\begin{equation*}
\phi_{\zeta,+1}\left( x\right) =\frac{1}{\sqrt{2}}\left( 
\begin{array}{c}
\psi^{(\zeta,+1)}\left( x\right) \\ 
\gamma^{0}\psi^{(\zeta,+1)}\left( x\right)%
\end{array}
\right) ,\;\;\phi_{\zeta,-1}\left( x\right) =\frac{1}{\sqrt{2}}\left( 
\begin{array}{c}
\psi^{(\zeta,-1)}\left( x\right) \\ 
\gamma^{0}\psi^{(\zeta,-1)}\left( x\right)%
\end{array}
\right) .
\end{equation*}
One can easily demonstrate that these two representations are physically
equivalent.

\subparagraph{\protect\large Acknowledgement}

S.P.G. and J.L.T. are grateful to FAPESP. D.M.G. acknowledges the support of
FAPESP, CNPq and DAAD.

\end{document}